\documentclass[12pt,a4paper,tightenlines,nofootinbib]{revtex4}
\usepackage{amssymb}
\usepackage{amsmath}
\usepackage{epsf}
\usepackage{psfrag}

\def\nablaslash{\not{\hbox{\kern-3pt $\nabla$}}}

\begin{document}

\author{Jaume Garriga$^{1}$ and Takahiro Tanaka$^{2}$}
\affiliation{$^1$ Departament de F{\'\i}sica Fonamental,
Universitat de Barcelona, Diagonal 647j, 08028 Barcelona, Spain}
\affiliation{$^2$ Department of Physics, Kyoto University, Kyoto, Japan}
\title{Can infrared gravitons screen $\Lambda$?}
\date{\today}

\begin{abstract}

It has been suggested that infrared gravitons in de Sitter space
may lead to a secular screening of the effective cosmological
constant.  This seems to clash with the naive expectation that the curvature
scalar should stay constant due to the Heisenberg equation of motion.
Here, we show that the tadpole correction to the local
expansion rate, which has been used in earlier analyses as an
indicator of a decaying effective $\Lambda$, is not gauge
invariant. On the other hand, we construct a gauge invariant operator which 
measures the renormalized curvature scalar smeared over an arbitrary window function, 
and we find that there is no secular screening of this quantity (to any given 
order in perturbation theory).

\end{abstract}

\maketitle

\section{Introduction}

In de Sitter space, long wavelength gravitons are frozen in,
producing a cumulative deformation of spacetime on large scales.
This can be seen, for instance, in the tree level graviton two
point function. In the transverse traceless gauge, the behaviour
of gravitons is similar to that of massless minimally coupled
scalars~\cite{staro}, and their two point function $\langle h(x)h(x')\rangle$
grows logarithmically with scale. Globally, such increasing
departure from a de Sitter metric cannot be undone by a gauge
transformation. Nevertheless, infrared gravitons do not contribute
to tidal forces on small scales. The tree level two point
function for the Riemann tensor $\langle R(x)R(x')\rangle$ is
infrared finite, and the contribution of gravitons with wavelength
much larger than the curvature scale $H^{-1}$ is in fact
negligible. Hence, to lowest order in perturbation theory, the
local geometry remains everywhere close to the unperturbed de
Sitter space \footnote{Provided, of course that $H$ is well below
the Planck scale. The graviton power spectrum is scale invariant
for wavelengths above $H^{-1}$, with amplitude of order $h\sim
H/M_p$.}.

It has long been suggested that graviton interactions may
dramatically alter this picture, potentially leading to infrared
screening of the cosmological constant~\cite{tw}. The basic idea
is the following. Gravitons carry energy and hence they are a
source of the gravitational field. Hence, it is conceivable that
the accumulation of infrared modes crossing the horizon in the
expanding de Sitter phase may backreact on the average expansion
rate of the universe. A priori, it is unclear whether infrared
gravitons can have much of an effect, since the ``energy'' in the
gravitational field is contained in derivatives of the metric. To
make a quantitative estimate, the authors of~\cite{tw} calculated
the graviton tadpole $\langle h_{\mu\nu}\rangle$ at the two loop
order. From that, they obtained the ``tadpole corrected'' expansion
rate of the universe $H(\langle h_{\mu\nu}\rangle )$, which turned
out to decrease quadratically with cosmic time. This slowing down
of the expansion rate of the ``averaged metric'', was interpreted as
a secular screening of $\Lambda$ by the long wavelength modes. If
true, this would be a spectacular effect of low energy quantum
gravity, with implications for the cosmological constant 
problem~\cite{wo04}.

The purpose of this paper is to reanalyze this problem, with an
emphasis on gauge invariance. In Section II we show
that the tadpole correction to the expansion rate, as defined in
Ref.~\cite{tw}, is not gauge invariant (and can in fact be given
an arbitrary time dependence). In Section III we discuss a
physically motivated gauge invariant definition of the expansion rate, which in the present context
essentially links it to the local value of the Ricci scalar. In  Section IV we calculate a gauge invariant
smeared expectation value of the Ricci scalar, suitably renormalized, showing that 
there is no infrared secular screening of this quantity. Our conclusions are summarized in Section IV.

\section{On the tadpole correction to the expansion rate}

The theory under consideration is pure gravity with a cosmological
constant. The action is given by
\begin{equation}
S_{gr} ={1\over 2 \kappa} \int \sqrt{-g} \left({\cal R} - 2
\Lambda\right) d^4 x~, \label{action}
\end{equation}
where ${\cal R}$ is the Ricci scalar and $\kappa = 8\pi G$. Here,
$G$ is Newton's constant. We are interested in perturbations
around the de Sitter space solution, and for definiteness we
shall adopt the flat chart description. The perturbed metric can
be written as
\begin{equation}
g_{\mu\nu}(x)= a^2(\eta)[\eta_{\mu\nu} + h_{\mu\nu}(x)]~.
\label{pertmet}
\end{equation}
Here $\eta_{\mu\nu}$ is the Minkowski metric, $a(\eta)=-1/(H_0\eta)$, with $-\infty <\eta<0$ the conformal time and $H_0$ the
constant unperturbed expansion rate.

To perform a systematic perturbative expansion using the path integral, 
we must add to~(\ref{action}) a gauge fixing term $S_{gf}=
-(1/2)\eta^{\mu\nu}F_{\mu}[h]F_{\nu}[h]$, where the function $F_\mu$ is such that $F_\mu[h]=0$ selects one representative out of a given
gauge orbit. In Ref.~\cite{tw} this function was chosen as
\begin{equation} F_{\mu}[h]\equiv a \left( h^\nu_{\mu,\nu} -
{1\over 2} h,_{\mu} + 2 h^\nu_\mu {a,_\nu \over a}\right)~.
\label{F}
\end{equation}
Here, indices are raised and lowered with the Minkowski metric
$\eta_{\mu\nu}$. Suitable counterterms will be
also be needed in order to remove divergences.\footnote{General
Relativity is non-renormalizable, and the number of counterterms
needed in $S_{ct}$ increases with the number of loops at which we calculate our observables. However,
the number is finite at any order, as it is usually the case in
effective field theories.} The total action takes the form
\begin{equation}
S_{tot} = S_{gr} + S_{gf} + S_{FP} + S_{ct}~,
\label{totalact}
\end{equation}
where $S_{FP}$ indicates the
Faddeev-Popov (FP) ghost terms and $S_{ct}$ the counterterms.
The graviton tadpole is defined by
\begin{equation}
\langle h_{\mu\nu} \rangle_F = \int_{CTP} {\cal D} \psi^{+}\ {\cal D}
 \psi^{-} 
\ h^{+}_{\mu\nu}\ e^{i S_{tot}[\psi^+]}\,  e^{-i S_{tot}[\psi^-]}~, \label{path}
\end{equation}
where the subindex $F$ refers to the gauge fixing 
function~(\ref{F}) and $\psi$ indicates the set of dynamical variables:
metric perturbations $h_{\mu\nu}$ and the FP ghosts and anti-ghosts. The
closed time path (CTP) version of the path integral is indicated, 
since we are interested in expectation values (rather than matrix
elements between in and out vacua).

The left hand side of Eq.~(\ref{path}) can be computed diagrammatically order by
order in perturbation theory. In the  gauge~(\ref{F}), the
propagator is infrared divergent in the limit of infinite 
volume~\cite{tw}, so it is convenient to compactify the spatial
directions and consider a finite (although in principle arbitrary
large) co-moving volume. If we choose a spatially homogeneous
initial state, symmetry requires that
\begin{equation}
\langle h_{\mu\nu} \rangle_F = A_F(\eta)\ \eta_{\mu\nu} +
B_F(\eta)\ t_{\mu} t_{\nu}~, \label{tadpole}
\end{equation}
where, $t^{\mu}=(\partial_\eta)^\mu$.

From~(\ref{pertmet}) and (\ref{tadpole}), the ``averaged'' metric
$\langle g_{\mu\nu}\rangle_F$ is a flat Friedmann-Robertson-Walker (FRW) metric, with expansion rate given
by
\begin{equation}
H_F\equiv H(\langle h_{\mu\nu}\rangle_F)= {d\ln[a
(1+A_F)^{1/2}]\over a\ (1+A_F-B_F)^{1/2} d \eta} = {H_0 \over
(1+A_F-B_F)^{1/2}}\left[ 1 - {1\over 2} {\eta A'_F\over (1+
A_F)}\right]~, \label{H}
\end{equation}
where the prime indicates derivative with respect to
$\eta$.

%The difference
%$$
%\Delta H_F \equiv \Delta H (\langle h_{\mu\nu}\rangle_F) \equiv
%H_F - H_0,
%$$
%will be refered to as the tadpole correction to the expansion
%rate.

In Ref.~\cite{tw}, Tsamis and Woodard calculated
$A_F$ and $B_F$ at the two loop order. Upon substitution 
in~(\ref{H}) they obtained
\begin{equation}
H_F= H_0 \left[1- 4\kappa^2\left({H_0\over 4 \pi}\right)^4
\left[{1\over 6} (H_0 t)^2 + {\cal O}(H_0 t)\right] + 0
(\kappa^6)\right]~, \label{decay}
\end{equation}
which decreases quadratically with cosmological time
$t=-H^{-1}\ln(H_0\eta)$. As mentioned in the introduction, this
result was interpreted in~\cite{tw} as a secular screening of the
effective cosmological constant by the infrared gravitons.
However, as we shall now discuss, $H_F$ is not
invariant under generic gauge transformations, and so the
above interpretation seems rather questionable.

Let us consider a new gauge fixing function $G[h]$ in the vicinity
of $F[h]$. If $F[h]=0$, then we can find a new metric perturbation
$$
\tilde h_{\mu\nu}= h_{\mu\nu}+\delta_\chi h_{\mu\nu}~,
$$
related to $h_{\mu\nu}$ by a gauge
transformation $\delta x^{\mu}=x'^\mu-x^\mu= \chi^{\mu}$,
such that $G[\tilde h]=0$.
Here
$$
\delta_\chi h_{\mu\nu}={2 a^{-2}} \nabla_{(\mu} \chi_{\nu)} +
{\cal O}(\chi^2)~,
$$
where $\nabla_\mu$ is the covariant derivative with respect to the
full metric $g_{\mu\nu}$ and $\chi_\nu=g_{\nu\lambda}
\chi^\lambda$.

Note that the gauge transformation will
in general depend on $h_{\mu\nu}$,
\begin{equation}
\chi^{\mu}=\chi^{\mu}[h]~, \label{gth}
\end{equation}
and even for simple changes of the gauge function $F[h]$ the
dependence of $\chi^{\mu}$ on $h$ can be quite non-trivial. The
point, however, is that for every $h$ this transformation will
exist. In what follows, we shall consider the class of gauge
function $G$ in the neighborhood of $F$ which are defined through
the equation
$$
F[h]=G[h + \delta_\chi h]~,
$$
for some $\chi$.

Both $S_{gr}$ and $S_{ct}$ are gauge invariant, $ S_{gr}[h] =
S_{gr}[h+\delta_\chi h]$,and $S_{ct}[h] = S_{ct}[h+\delta_\chi
h]$. Moreover $(S_{gf})_F [h]=(S_{gf})_G [h+\delta_\chi h]$ and
$(S_{FP})_F [h]=(S_{FP})_G [h+\delta_\chi h].$ It is then
straightforward to show, by changing variables in~(\ref{path}),
that
\begin{equation}
\langle h_{\mu\nu} \rangle_G = \langle h_{\mu\nu}+\delta_\chi
h_{\mu\nu} \rangle_F~.
\end{equation}
The variation of the tadpole under gauge transformations is thus
given by
\begin{eqnarray}
&& \langle h_{\mu\nu} \rangle_G - \langle h_{\mu\nu} \rangle_F
=\langle\delta_\chi h_{\mu\nu} \rangle\equiv (\delta_\chi A)\
\eta_{\mu\nu}+ (\delta_\chi B)\ t_{\mu}t_\nu 
\cr  &&\quad =
\langle [\eta_{\lambda\nu}+h_{\lambda\nu}] \chi^\lambda_{,\mu} +
[\eta_{\lambda\mu}+h_{\lambda\mu}] \chi^\lambda_{,\nu} 
+h_{\mu\nu,\lambda} \chi^{\lambda}-{2\over \eta}
[\eta_{\mu\nu}+h_{\mu\nu}] \chi^0   \rangle + O(\chi^2)~.
\label{gt}
\end{eqnarray}
Here $\delta_\chi A$ and $\delta_\chi B$ represent the changes in
$A(\eta)$ and $B(\eta)$ defined in~(\ref{tadpole}).
In the above expression $\chi$ is treated as a small quantity, but
$h_{\mu\nu}$ is not necessarily small.

In the case when $\chi^\mu$ is a c-number (and by this we mean a function independent of $h$), 
the only transformation
compatible with the symmetries of a flat FRW is
\begin{equation}
\chi^\mu = f(\eta) t^\mu~. \label{chi0}
\end{equation}
If we neglect $h_{\mu\nu}$ in Eq.~(\ref{gt}),
the expansion rate $H$ is invariant under gauge
transformations. The reason is that for a flat FRW the expansion
rate is given in terms of the temporal component of the Einstein
tensor
\begin{equation}
H^2={1\over 3} G^0_0~. \label{HG}
\end{equation}
The background is such that
\begin{equation}
G^\mu_\nu - \Lambda\ \delta^\mu_\nu=0~. \label{eom}
\end{equation}
It follows that $\delta_\chi G^\mu_\nu =0$, and from~(\ref{HG})
$\delta_\chi H = 0$. More explicitly, from~(\ref{gt}) we have
$\delta_\chi A = -(2f/\eta)$ and $\delta_\chi B = -2 f'$, and
linearizing~(\ref{H}) we have \footnote{A derivation of~(\ref{gi})
under similar assumptions was given in~\cite{tw}. This, however,
does not establish that $H(\langle h_{\mu\nu} \rangle)$ will be
invariant. As shown below, Eq.~(\ref{gi}) does not hold for
generic gauge transformations.
%There,
%in the expansion of $\delta_\chi H$ in powers
%of $\chi$, the zero order value of $A$ and $B$ was neglected. This is consistent %with neglecting $h_{\mu\nu}$ in~(\ref{gt}). Also, the change in the gauge %function $F$ was taken to be a c-number, and the calculation was carried out at %the lowest order in perturbation theory. This is consistent with taking %$\chi^\mu$ to be a c-number, independent of $h$.
}
\begin{equation}
\delta_\chi H \equiv H_G-H_F = (H_0/2) [\delta_\chi B-\delta_\chi
A-\eta (\delta_\chi A)'] = 0~. \label{gi}
\end{equation}
Provided that $\chi^{\mu}$ is independent of $h$, the above
consideration can be extended to the case when the tadpole is
non-vanishing. Using~(\ref{tadpole}) in~(\ref{gt}) we find
$\delta_\chi A = - (2 f/\eta) (1 + A) + A'f$ and $\delta_\chi B =
-2f'(1 + A - B) - (2f/\eta) B + B' f$. Substituting these
variations in~(\ref{H}), it is straightforward to check that
\begin{equation}
\delta_\chi H(\eta) = {dH(\eta)\over d\eta} f(\eta) + \cdots~.
\label{Hp}
\end{equation}
Here, $H(\eta)$ represents the right hand side of~(\ref{H}), which
depends on time through $A$ and $B$, and the ellipsis denote
higher orders in $\chi$.

The simple form of~(\ref{Hp}) is easily
understood. To lowest order in $\chi$, the variation $\delta_\chi
h_{\mu\nu}$ is {\em exactly} linear in $h_{\mu\nu}$ [the metric perturbation $h_{\mu\nu}$ is {\em not} treated as a small parameter in
Eq.~(\ref{gt})]. Because of that, $\langle h_{\mu\nu}\rangle$
transforms like a classical metric under infinitesimal c-number
gauge transformations $\chi^{\mu}$,
\begin{equation}
\langle\delta_\chi h_{\mu\nu} \rangle = \delta_\chi \langle
h_{\mu\nu} \rangle + \cdots~, \label{diff}
\end{equation}
Eq.~(\ref{Hp}) follows immediately by noting that $H^2= (1/3)\
G^0_0$ is a scalar under redefinitions of the time coordinate
(because it has mixed temporal indices). The gauge 
dependence~(\ref{Hp}) is therefore rather irrelevant: 
it indicates that we
have changed the parametrization of a time dependent function, but
this does not change the value of the expansion rate $H$ as a
function of proper time $t$ [defined by $dt= a
(1+A-B)^{1/2}d\eta$].

However, the above conclusions do not apply to the case where the
gauge transformations depend on the metric~\cite{unruh98}
\footnote{For illustration, note that even the simplest changes in the gauge function $F$ may
lead to a complicated dependence of $\chi$ on $h$~(\ref{genga}).
Consider for instance the one parameter class of gauges
\begin{equation}
\label{class} F^{(\alpha)}_\mu =F_{\mu}[h]\equiv a \left(
h^\nu_{\mu,\nu} + \alpha h,_{\mu} + 2 h^\nu_\mu {a,_\nu \over
a}\right)~,
\end{equation}
out of which~(\ref{F}) corresponds to $\alpha=-1/2$. A change of
$\alpha$ corresponds to $\delta F_\mu = a \ h,_\mu \delta \alpha$,
and we have
$$
\int d x' {\delta F_\mu[h] \over \delta h_{\rho\sigma}(x') }\ \delta_\chi
h_{\rho\sigma}(x') = a \ h,_\mu \delta \alpha~.
$$
By introducing the explicit expression of $F$ and $\delta_\chi
h_{\mu\nu}$, this equation takes the form
\begin{equation}
{\cal O}_{\mu\nu}[h] \chi^\nu = h,_\mu~, \label{partdiff}
\end{equation}
where ${\cal O}_{\mu\nu}[h]$ is a second order differential
operator whose coefficients depend on $h_{\mu\nu}$ and its first
and second derivatives. Eq.~(\ref{partdiff}) should in principle
be solved in order to find $\chi$ in terms of $h$. It is clear
that in this case the dependence will be highly non-local (and
difficult to find explicitly), but the point is that we cannot
restrict consideration to c-number gauge transformations, since
generically $\chi$ depends on $h$.}
\begin{equation}
\chi^{\mu} = \chi^{\mu}[h]~.
 \label{genga}
\end{equation}
%Consider, for instance, a gauge function of the form $\chi^\mu =
%\sum f^{(i)}(\eta) \chi^{(i)\mu}[h]$, where $\chi^{(i)}$ is a set
%of independent functions of the metric perturbation $h$ and its
%derivatives (e.g. a set of polynomials), and $f^{(i)}$ are
%arbitrary functions of time. The vector $t_\mu$, as well as
%arbitrary functions of time, can also be used in constructing each
%of the individual terms $\chi^{(i)}$.
For generic choices of
$\chi[h]$, we should expect $$\langle h \chi \rangle \neq \langle h \rangle
\langle \chi \rangle~,$$and from Eq.~(\ref{gt}), we should likewise expect that
$$
\langle\delta_\chi h_{\mu\nu} \rangle \neq
\delta_{\langle\chi\rangle} \langle h_{\mu\nu} \rangle + \cdots~.
$$
We may thus anticipate that, in general,
the expectation value of the gauge transformed metric perturbation $\langle h +
\delta_\chi h\rangle$ will not be gauge equivalent to the original
one $\langle h \rangle$.

The expansion rate $H(\langle h_{\mu\nu} \rangle)$ as a function
of proper time $t$ would be gauge invariant (and therefore
meaningful) if and only if $\langle h + \delta_\chi h\rangle$ is
related to $\langle h \rangle$ by a time reparametrization (see e.g. the
discussion around Eq.~(\ref{diff})). In equations, this means that for
each $\chi^\mu[h]$ we should be able to find a vector $\xi^\mu$ such 
that
\begin{equation}
\langle\delta_\chi h_{\mu\nu} \rangle = \delta_{\xi} \langle
h_{\mu\nu} \rangle + \cdots~, \label{diff2}
\end{equation}
where, by symmetry, $\xi^{\mu}=g(\eta)t^{\mu}$. However, an equation
of this sort cannot hold for a generic $\chi[h]$. To illustrate
the point, we may restrict ourselves to {\em lowest order} in
perturbation theory, where the expectation value of odd functions
of $h$ will vanish. Let us therefore take $\chi$ to be an odd
function of the metric perturbation $h$. In this case 
Eq.~(\ref{diff2}) to lowest order in perturbation theory reads
\begin{equation}
\langle 2 h_{\lambda(\mu}\chi^\lambda_{,\nu)} +
h_{\mu\nu,\lambda}\chi^\lambda -{2\over \eta}
h_{\mu\nu}\chi^0\rangle = -{2g\over\eta}\eta_{\mu\nu} -2 g' t_\mu
t_\nu~. 
\label{impossible}
\end{equation}
Here we have used $\xi^{\mu}=g(\eta)t^{\mu}$, as dictated by
symmetry. When we consider rescaling given by 
$\chi^\mu[h] \to k(\eta)\chi^\mu[h]$ with an arbitrary function $k$,  
we have the terms with $k'(\eta)$ and those with $k(\eta)$ on 
the left hand side of Eq.~(\ref{impossible}). 
In order to satisfy this equality for an arbitrary function $k$, 
the right hand side also has to have the terms with 
$k'(\eta)$ and those with $k(\eta)$. This requires that $g(\eta)$ 
should also transform to $k(\eta)g(\eta)$.  
Comparing the term proportional to $k'(\eta)$, 
we immediately find 
\begin{equation}
g=\langle h_{0\lambda} \chi^\lambda \rangle~.
\end{equation}
Then, if we choose $\chi^\mu = \chi^0[h] t^\mu$, where $\chi^0$ is
an arbitrary odd function of $h_{\mu\nu}$ (including possible
arbitrary explicit dependence on $\eta$ and contraction of
internal indices with the vector $t^\mu$), the $\mu=\nu=0$
component of~(\ref{impossible}) reads
\begin{equation}
\langle h_{00,0}\ \chi^0 \rangle = 0~. \label{no}
\end{equation}
Clearly, this equation does not hold in generic gauges and for
generic choices of $\chi^0$. In particular, it does not hold in
the gauge defined in Eq.~(\ref{F}), which completes our proof
\footnote{Eq.~(\ref{no}) may accidentally hold in some gauge for
all possible functions $\chi^0$. For instance, it holds in the
transverse and traceless gauge, since $\langle h_{00}\rangle$
vanishes to lowest order. In that case we should examine the other
components of Eq.~(\ref{impossible}) to see whether they can hold
for generic $\chi$.}.

It follows that the ``tadpole corrected'' expansion rate is not
physically meaningful. Rather, given the enormous freedom in
choosing $\chi^\mu$ (which can include arbitrary functions of $h$
and $\eta$) it appears that $H(\langle h_{\mu\nu} \rangle)$ can in
fact be given arbitrary dependence on proper time $t$.

\section{Observables}

Gravitational radiation of wavelength shorter than the Hubble
radius has an impact on the background expansion rate. Formally,
this can be accounted for in the so-called Isaacson approximation
(see e.g.~\cite{MTW}), where the Einstein tensor is split into a
background ``long wavelength'' contribution and the contribution
from short wavelength gravitational waves. A classical bath of
short wavelength gravity waves does modify the time evolution of
the scale factor, much like a bath of radiation would. However,
this does not mean that it screens the cosmological constant,
which also contributes to the expansion rate as usual.

On the other hand, here we are not interested in the effect of
short wavelength modes but in the collective effect of infrared
graviton modes and their interactions. Could these cause a secular
screening of the cosmological constant? What we mean by this is 
an adiabatic erosion of the expansion rate, such as the one suggested 
by Eq.~(\ref{decay}), which would lead to an initial quasi-de Sitter 
phase with 
\begin{equation}
|\dot H| \ll H^2~.\label{qds}
\end{equation}

As emphasized in the previous subsection, instead of calculating the expectation
value of the metric, it is important to look for some gauge invariant characterization of the expansion
rate. One such observable was suggested by Abramo and 
Woodard~\cite{Abramo:2001db}. The basic idea is to consider the value of a scalar
field $\varphi$ conformally 
coupled to gravity, with a constant
source term $J$:
\begin{equation} \left[\Box + {{\cal R} \over 6}\right] \phi = J~.
\label{source}
\end{equation} 
In a flat FRW universe, the Ricci scalar is given by
\begin{equation}
{\cal R}= 12 H^2(t) + 6 \dot H(t)~.\label{riccifrw}
\end{equation}
For instance, when the scale factor takes the form $a \propto
t^p$, we have $H(t)= p t^{-1}$ and
$
{\cal R}= (12p^2-6 p) t^{-2}.
$
The general solution of Eq.~(\ref{source}) in this case is of the form \begin{equation}
\phi(t) = A_+ t^{\alpha_+} + A_- t^{\alpha_-} +  \left({2 p^2
\over 2 p^2 + 5 p + 2}\right) {J\over 2 H^2(t)}~,\label{soln}
\end{equation}
where $A_\pm$ are arbitrary constants and
$$
\alpha_\pm = {-3 p \pm \sqrt{p^2+4p} \over 2}~.
$$
The last term in Eq.~(\ref{source}) is proportional to $t^2$. 
As we have $\Re \alpha_\pm < 2$ for $p>-4/3$, the two first terms in~(\ref{soln}) decay faster than the last term. 
Therefore at late times only the third term will be
important. The limit of quasi de Sitter expansion ($\dot H \ll
H^2$) corresponds to $p\gg 1$, and in this case we have
\begin{equation}
\phi(t) \approx {J\over 2 H^2(t)}\quad (t\to \infty, p\gg 1)~.
\label{result}
\end{equation}
We may interpret this result in the following way. The source $J$
creates the field $\phi$. In turn, the field is diluted by the
expansion, which causes the amplitude to fall off as the inverse
of the scale factor. At late times, the value of $\phi$ is
dominated by the field created by the source $J$ during the last
expansion time, on the surface of a sphere of radius $H^{-1}$,
while the initial conditions set by the coefficients $A_\pm$
become irrelevant. Thus, the late time asymptotic behaviour of
$\phi(t)$ sourced by a constant $J$ is a measure of the time
dependent expansion rate $H(t)$, through Eq.~(\ref{result}).
Let us now consider the quantum theory. To avoid the introduction
of a new quantum scalar field in the system, it was suggested in
Ref.~\cite{Abramo:2001db} 
that $\phi(t)$ be defined as the expectation value
of inverse of the retarded conformal propagator acting on the
constant source
\begin{equation}
\phi(t) = \langle\ {1 \over \Box +{1\over 6} {\cal R}}\ J \
\rangle~.\label{awo}
\end{equation}
This quantity would then be a characterization of the inverse 
of the square of the expansion rate, through the
identification~(\ref{result}). Although this seems to be
an appropriate definition, it is certainly somewhat
complicated because of the non-local character of the
operator within brackets.

In the present context, however, there is a much
simpler alternative which is equally well motivated. Rather than 
keeping a constant source and looking for
the field it creates at a given point, we may ask the converse:
What source $J(x)$ will give a constant field $\phi(t)\to 1$ at
late times? Classically, if the expansion rate is a constant,
$H=H_0$, then a constant source will produce an asymptotically
constant field $\phi(t)\to J/(2 H_0^2)$ at $t\to \infty$. In the
quasi-de Sitter limit $(p\gg 1)$, a constant field $\phi\to 1$ is
caused by a source of the form $J \to 2 H^2(t)$. In this sense,
the source which is needed in order to keep $\phi\to 1$ can be
used as a measure of the local expansion rate. Needless to say,
this measure reduces trivially to the curvature scalar
\begin{equation}
2 H^2(t) \approx J(t) = \left({\Box +{1\over 6} {\cal R}}\right) 1
={1\over 6} {\cal R}~.
\label{JandR}
\end{equation}
This illustrates the fact that in a quasi-de Sitter phase 
we may adopt $\sqrt{{\cal R}/12}$ as a {\em local} definition of the expansion
rate, because the second term in~(\ref{riccifrw}) is negligible.
The expansion rate can change by a large amount in the course of time,
but as long as it does so adiabatically, the curvature scalar will be a good
tracer of $H(t)$.

Now, in the quantum theory, the metric fluctuates. If we adopt the
expectation value of the classical expression as our
definition of $H(t)$, we have
\begin{equation}
H^2(t) \approx {1\over 12}\ \langle{\cal
R}\rangle~.\label{gidh}
\end{equation}
Naively, in pure gravity with a cosmological constant, we may expect to have a relation of the form
\begin{equation} 
\langle{\cal R}(x)\rangle = 4\Lambda~,
\label{lt}
\end{equation}
which would readily imply the constancy of $H(t)$, with no room for a secular screening.
Intuitively, Eq.~(\ref{lt}) seems to follow from the Heisenberg equation of motion, 
which the field operator is supposed to satisfy identically . Nevertheless, the definition of
a physically meaningful $\langle{\cal R}\rangle$, and the proof of an equation of the form~(\ref{lt}), involves a number of subtleties related to gauge invariance
and renormalization (see also~\cite{Losic:2006ht,Hollands:2004yh}). A full discussion of this
point is postponed to the next Section.

Before closing, one comment is in order about the 
classical back-reaction. 
Classically, in a flat FRW universe, we have 
\begin{equation}
{\dot H \over H^2}= -{3\over 2} (1+w_{\rm eff})~,
\end{equation}
where $w_{\rm eff}=p/\rho$ is 
the ratio of pressure $p$ to energy 
density $\rho$. As mentioned at the
beginning of this section, the classical back-reaction due to
short-wavelength gravitational waves modifies the expansion law 
like a usual radiation field, with $w_{\rm rad}=1/3$. If the density in radiation is 
comparable to the cosmological term, then $(1+w_{\rm eff})$ will not be
small, and from~(\ref{riccifrw}), ${\cal R}$ will not be a good tracer
of $H(t)$. At the classical level, 
${\cal R}$ is constant, and hence it is completely insensitive 
to the classical back-reaction effect. In this respect, the observable 
originally proposed 
%by Abramo and Woodard 
in Ref.~\cite{Abramo:2001db}, given in Eq.~(\ref{awo}),
cannot sense the traceless component of the energy momentum tensor, either. 
Clearly, $\phi=$const. is one of the solutions when we assume $J=$const. 
Although there are various solutions for $\phi$ for a constant $J$, 
this variety is due to the degrees of freedom of the initial conditions,
which are supposed to be irrelevant at late times. 
The study of alternative observables which are, at least, sensitive to the 
back-reaction effect caused by classical gravitational waves, is postponed 
for future work.

\section{Is there a secular screening?}

As discussed above, the Ricci scalar ${\cal R}$ is a good indicator of the adiabatic evolution of 
the expansion rate in a quasi-de Sitter phase. Here, we will discuss the calculation of its 
renormalized expectation value in the theory of pure gravity.  
For this purpose, it will be quite important to work with quantities which are invariant under
diffeomorphisms. When we consider the gauge transformation 
\begin{equation}
x^\mu\to \bar x^\mu=x^\mu -\chi^\mu~, 
\label{ctrans}
\end{equation}
${\cal R}(x)$ 
transforms like 
${\cal R}(x)\to {\cal \bar R}(x)\approx 
{\cal R}(x)+{\cal R}_{,\mu}\chi^\mu$.
In this sense ${\cal R}$ is not invariant. However, we do not really need to 
measure the value of curvature at a specified point in spacetime. Rather, it will 
be sufficient for our purposes
to consider its value smeared over a certain sample volume. Let us introduce a window function 
$W(x)$ which by definition transforms as a spacetime scalar. 
Then, for any spacetime scalar operator ${\cal O}(x)$, the integral
$\int d^4 x\, \sqrt{-g}\ W(x) {\cal O}(x)$ is manifestly gauge
invariant.\footnote{
\label{footnote6}
Note that
$W(x)=W_z(x)\equiv \delta^{(4)}(x^\mu-z^\mu)/\sqrt{-g(x)}$, is {\em not} a suitable window
function. Although this is a scalar with respect to the transformation of $x$, 
it is also so with respect to the transformation of $z$. The bi-scalar transforms as 
$\delta W_z(x)=\chi^\mu(x){(\partial W_z(x)/ \partial x^\mu)}+
                  \chi^\mu(z){(\partial W_z(x)/ \partial z^\mu)}, $
and for that reason $\int d^4x\,\sqrt{-g}\,{\cal R}(x) W_z(x)={\cal R}(z)$
is not gauge invariant (in agreement with the discussion below Eq.~(\ref{ctrans})). }
It will be useful to introduce the following notation for the expectation value of this
quantity:
$$\left\langle {\cal O} \right\rangle_W\equiv 
  \left\langle {\rm phys}\left\vert \int d^4 x\, \sqrt{-g}\ W(x) 
{\cal O}(x) \right\vert {\rm phys} \right\rangle~.$$ 
A precise definition of the arbitrary physical state $\vert {\rm phys}\rangle$ will be given 
below.

The basic goal of this Section is to show that the equation
\begin{equation} 
\langle {\cal R}_{\rm ren}(x)\rangle_W
= 4\Lambda\ \langle 1_{\rm ren} \rangle_W~,
\label{lt2}
\end{equation}
holds for any scalar window function $W$. The definition of the operators ${\cal R}_{\rm ren}(x)$ and $1_{\rm ren}$
appearing in~(\ref{lt2}) requires explanation. In the path integral approach, we can calculate the
$n$-point functions $\langle h_{\mu\nu}(x_1)\cdots h_{\mu\nu}(x_n)\rangle$ to arbitrary loop order. All divergences 
in this calculation can be reabsorbed by diagrams involving the vertices generated by the 
counterterms in $S_{ct}$. Still, such renormalized $n$-point functions will not be free from divergences 
in the coincidence limit, when two or more of the $n$ points are brought to sit on top of each other. 
Since $\sqrt{-g}{\cal R}$ contains the coincidence limit of $n$-point functions of 
$h_{\mu\nu}$, the counterterms in $S_{ct}$ will fail to render a finite expectation value for 
$\sqrt{-g}{\cal R}$.  This situation, of course, is not specific to gravity, and the problem is remedied once we
introduce a probe field which couples to the composite operator of interest. This will allow us to define a 
suitable regularized operator $\sqrt{-g}{\cal R}_{\rm ren}$ whose renormalized expectation value is finite.

It is instructive to start by considering the simpler example of an interacting scalar field $\psi$ 
in Minkowski spacetime~\cite{Vilenkin:1982wt}. In this case, the 
two point function $\langle \psi(x)\psi(x')\rangle$ 
is finite after 
renormalization, but its coincidence limit $\langle \psi^2(x)\rangle$ 
is still divergent. 
In the context of a single free scalar field, there is no counterterm 
to renormalize the value of $\langle \psi^2(x)\rangle$. On the other hand,
we can only measure this seemingly divergent quantity through some interaction. 
Let us therefore introduce a coupling to a probe scalar field $\phi$ via the 
interaction Lagrangian $-\lambda\psi^2\phi^2/2$. Now, we can ``measure'' 
$\lambda \langle \psi^2(x)\rangle$ as a contribution to the mass 
of the probe field $\phi$. Here, the probe field is treated as classical,
meaning that we neglect all the loop diagrams containing its propagator. 
We also assume that its amplitude is infinitesimally small. 
Now, $g \langle \psi^2(x)\rangle$ can be renormalized because the divergence in $\langle \psi^2(x)\rangle$
can be absorbed by a mass counterterm of the $\phi$-field. Hence, we have found a regularized
operator $\lambda \psi_{\rm ren}^2(x) = \lambda \psi^2(x) + \delta m^2_\phi$ 
whose renormalized expectation value $$m^2_{\phi({\rm ren})}\equiv\langle\lambda \psi_{\rm ren}^2(x)\rangle~,$$ 
is finite by virtue of the probe field counterterm $\delta{\cal L}_\phi= -(\delta m^2_\phi)\phi^2/2$. 

The same argument works for 
$\sqrt{-g}{\cal R}$.  
We consider a probe massless scalar field $\phi$ with 
the curvature coupling as we discussed in the preceding section. 
The action we add is 
\begin{equation}
 S^{\phi}+S_{ct}^{\phi}=-{1\over 2}\int d^4x\sqrt{-g}\left(
    g^{\mu\nu} (\partial_\mu \phi) (\partial_\nu \phi)
    +{\xi }{\cal R}\phi^2 + \delta {\cal M}^2_{\phi}[g] \phi^2
     \right)~, 
\end{equation}
where the mass counterterm $\delta {\cal M}^2_{\phi}[g]$ is made up of curvature 
invariants, and will be further specified below.
We may now define 
\begin{equation}
\xi {\cal R}^{\rm ren}\ \equiv {\xi }{\cal R}+ \delta {\cal M}^2_{\phi}~,
\label{rnm1}
\end{equation}
whose renormalized expectation value may be thought of as the local value of 
the mass of the $\phi$ field
\begin{equation}
m^2_{\phi \rm (ren)}[W]\equiv {\left\langle 
\xi{\cal R}^{\rm ren}(x)\right\rangle_W \over \langle 1^{\rm ren} \rangle_W}~.
\label{rnm2}
\end{equation}
Since the volume $\langle 1\rangle_W=
  \left\langle {\rm phys}\left\vert \int d^4 x\, \sqrt{-g}\ W(x) 
 \right\vert {\rm phys} \right\rangle$ contains polynomials of 
$h_{\mu\nu}$ through $\sqrt{-g}$, this quantity is also 
divergent. Hence, we have to renormalize this expression by adding 
a counterterm $\delta_{\rm vol}$, i.e. 
$\langle 1^{\rm ren}\rangle_W=\langle 1+\delta_{\rm vol}\rangle_W$. 
To be more precise, in order to renormalize the volume, 
we need to add another probe field to measure it. 
For example, we can consider a scalar field with $\xi=0$ as a probe. 
In this case the renormalized volume integral of its mass will 
measure the renormalized volume. 
%Notice that, since we evaluate the 
%volume integral of the right hand side of Eq.~(\ref{JandR}), 
%the left hand side of should also to be understood as a volume integral of
%$J$. Hence, it will be natural to consider that 
%Eq.~(\ref{remain}) devided by the volume $\langle 1^{\rm ren}\rangle_W$ is 
%the averaged value of $J$ for the window function $W(x)$. 
 
The value of 
$m^2_{\phi({\rm ren})}[W]$ will 
change depending on the choice of the finite part of counterterms. 
However, if there is a choice of counterterms in which the relation 
$m^2_{\phi \rm (ren)}[W] =  4\xi\Lambda $ 
is maintained independently of the window function $W$, it is such renormalization conditions 
that are natural and appropriate for the theory that we are considering.  
Finite renormalization of local counterterms will correspond to introducing new 
interactions between the probe field and gravity, different from 
the original curvature coupling term.  We shall not pursue the consideration of such interactions here. 
They correspond to higher order irrelevant operators (which are {\em not} expected to lead to infrared effects
of the sort we are interested in). Thus, the basic question is whether we can choose local counterterms 
which make the renormalized value of $m^2_{\phi({\rm ren})}$ to be constant.  

The key identity is 
\begin{eqnarray}
0& = & -i \int_{\phi=0} {\cal D}\psi \int d^4 x {\delta\over \delta
 \tilde h_{\mu\nu}(x)}
    W(x) g_{\mu\nu}(x)e^{iS_{tot}}\cr
 & = & \left\langle
   {1 \over 2\kappa}({\cal R}-4\Lambda)
  +{g_{\mu\nu}\over \sqrt{-g}}
   {\delta\left(S_{gf+FP}+S_{ct}\right)
      \over \delta \tilde h_{\mu\nu}} 
      \right\rangle_{W}^{(\phi=0)}~, 
\label{phi=0}
\end{eqnarray}
where we assumed that the window function $W(x)$ is independent of
$\tilde h_{\mu\nu}\equiv g_{\mu\nu}-g^{(0)}_{\mu\nu}$. 
{$g^{(0)}_{\mu\nu}$ is the background metric which can be different
from the de Sitter one as long as it solves the Einstein equations.} 
The first equality follows from functional 
integration
by parts. In the second equality, we have dropped the term $-16i\delta(0)$ which arises from functional 
differentiation
${\delta g_{\mu\nu}(x)/\delta \tilde h_{\mu\nu}(x)}$. It is clear that this naively divergent term is
local, and can be grouped together with $S_{ct}$. In fact, such functional derivative vanishes in
dimensional regularization. All the variables in the path integral are 
to be understood as $(+)$-fields and the integral over $(-)$-fields has been
abbreviated, since the
CTP formalism is not essential for the current discussion. 

We submit that
the appropriate choice of $\delta{\cal M}^2_{\phi}$ which
implements our renormalization 
condition is given by
\begin{equation}
\delta{\cal M}^2_{\phi}
={2\kappa\xi} {g_{\mu\nu}\over \sqrt{-g}}
   {\delta S_{ct}\over \delta \tilde h_{\mu\nu}} +
  4\xi \Lambda \delta_{\rm vol}~.
\end{equation}
Here, $S_{ct}$ is the counterterm action for the theory of pure gravity,
without the $\phi$ field. 
Thus,  $\delta{\cal M}^2_{\phi}$ is local as long as $S_{ct}$ is so. 
The other part 
$4\Lambda \delta_{\rm vol}$ is also local. 
Then, substituting in Eq.~$(\ref{rnm1})$, we have
\begin{equation}
\langle {\cal R}^{\rm ren} \rangle_W=4\Lambda\ \langle 1^{\rm ren}\rangle_W\ 
   -2\kappa \left\langle
  {g_{\mu\nu}\over \sqrt{-g}}
   {\delta S_{gf+FP}
      \over \delta \tilde h_{\mu\nu}} 
       \right\rangle_W^{(\phi=0)}~, \label{remain}
\end{equation}
where we have used Eq.~(\ref{phi=0}).

The remaining task is to show that the second term in the right hand side of~(\ref{remain}) vanishes 
when the expectation value is taken for physical states. 
To show this, the essential point is to understand what is
meant by the physical state. Since $\phi$ is set to $0$, we neglect it completely in the following
discussion. It will be very convenient for our purposes to follow the standard construction 
for gauge fixing based on the BRST invariance~\cite{kugo}. 
The gauge transformation changes 
$\tilde h_{\mu\nu}(x)\to 
\bar {\tilde h}_{\mu\nu}(x)=\tilde h_{\mu\nu}(x)+\delta \tilde h_{\mu\nu}(x)$ with   
\begin{equation}
 \delta \tilde h_{\mu\nu}(x)=g_{\mu\rho}\chi^\rho_{~,\nu} 
  +g_{\nu\rho}\chi^\rho_{~,\mu} +g_{\mu\nu,\rho}\chi^\rho~.
\label{eqAP2}
\end{equation}
The BRST transformation $\delta_B$ of $\tilde h_{\mu\nu}$ is obtained by 
simply replacing $\chi^\mu$ with 
a Grassmanian field $c^\mu$ in Eq.~(\ref{eqAP2}). The BRST transformation of $c^\mu$ is 
determined by requiring the nilpotency of the BRST transformation, 
$\delta_B^2 \tilde h_{\mu\nu}=0$. Different from the usual 
gauge theory, this equation does not determine $\delta_B c^{\mu}$
locally. The obtained equation contains derivatives of $\delta_B
c^{\mu}$. 
Hence, we do not give an explicit expression for $\delta_B c^{\mu}$, 
which is not required below. 
We add the anti-ghost field $\bar c^\mu$ and its BRST transformation
introduces 
$B^\mu$-field as $\delta_B \bar c^\mu =i B^\mu$. From the requirement of 
nilpotency of the BRST transformation, we have $\delta_B B^\mu=0$. 
After these preparations, for an arbitrary gauge fixing function
$F_\mu[\tilde h_{\alpha\beta}]$,  
the gauge fixing term and the Faddeev-Popov ghost term are
simultaneously given by 
\begin{eqnarray*}
 S_{gf+FP}=\int d^4x\, {\cal L}_{gf+FP}~,
\end{eqnarray*}
with
\begin{eqnarray}
{\cal L}_{gf+FP} & = & -i\delta_B\left[
    \bar c^\mu\left(
    F_{\mu}+{1\over 2}\alpha B_\mu\right)\right]\cr
   & = & B^\mu \left(
    F_{\mu}+{1\over 2}\alpha B_\mu\right)
    -i (\delta_B \tilde h_{\alpha\beta}) {\delta F_{\mu}\over \delta
    \tilde h_{\alpha\beta}} 
     \bar c^\mu~,   
\end{eqnarray}
where indices in ${\cal L}_{gf+FP}$ 
are raised and lowered by using the background 
metric $g^{(0)}_{\mu\nu}$. 
Since $F_\mu[\tilde h_{\alpha\beta}]$ may contain differentiation of 
$\tilde h_{\alpha\beta}$, 
${\delta F_{\mu}/\delta \tilde h_{\alpha\beta}}$ is understood 
as the derivative operator
that is obtained by the usual variational principle.
In the present case it acts on $\bar c^\mu$.   
For simplicity, we assume that $F_\mu[\tilde h_{\alpha\beta}]$ is linear
in $\tilde h_{\alpha\beta}$. 
Hence, ${\delta F_{\mu}/\delta \tilde h_{\alpha\beta}}$ is an operator solely written in 
terms of the background quantities. 

Let us now consider the physical observables and the physical states. In the BRST formalism, 
observables are BRST invariant quantities. This corresponds to the usual notion of gauge invariant 
variables such as the Bardeen parameter at the linear order. 
We should note that $\delta_B s(x)\ne 0$ for a scalar $s(x)$, 
which is the reason why we had to introduce a window function $W(x)$ to evaluate the expectation value of
${\cal R}$. An observable ${\cal O}$ satisfies $[Q_B,{\cal O}\}=0$, where $Q_B$ is the BRST charge defined 
in such a way that $\delta_B *=[Q_B,*\}$. Correspondingly, physical states are also required to be BRST
invariant. Hence, they must satisfy 
$$Q_B\vert {\rm phys}\rangle=0~.$$
Therefore, for physical states, any operator that can be written in
the ``exact'' form $[Q_B, *\}$ has vanishing expectation value. 

Then, using $\delta_BW(x)=c^\mu\partial_\mu W(x)$, which is the standard transformation rule
for any scalar quantity, it is straightforward to show that 
\begin{eqnarray}
&& \int d^4x\, W 
  g_{\mu\nu} {\delta \over \delta \tilde h_{\mu\nu}}
 \int d^4x'\,{\cal L}_{gf+FP}\cr
&&\qquad
 = \int d^4x\, W  
   \left(g_{\mu\nu} 
   {\delta F_{\alpha}\over \delta \tilde h_{\mu\nu}} B^\alpha
  -i (\delta_B \tilde h_{\rho\sigma}) 
     {\delta F_{\alpha}\over \delta \tilde h_{\rho\sigma}}\bar c^\alpha
  +i \partial_\mu \left(c^\mu g_{\rho\sigma}
     {\delta F_{\alpha}\over \delta \tilde h_{\rho\sigma}}\bar c^\alpha
    \right)\right) \cr
&&\qquad
 = 
 \left[Q_B, 
   -i\int d^4x\,W{g_{\mu\nu}} 
  {\delta F_{\alpha}\over \delta \tilde h_{\mu\nu}}\bar c^\alpha
   \right\}~. 
\end{eqnarray}
Hence, the contribution from $S_{gf+FP}$ vanishes when  
the expectation value is calculated for physical states. 
This finally establishes our claim 
that we can choose local counterterms such that
$ \langle {\cal R}_{\rm ren}(x)\rangle_W
= 4\Lambda\ \langle 1_{\rm ren} \rangle_W \, $
holds for an arbitrary scalar window function $W(x)$. This simply means
that ${\cal R}_{\rm ren}(x)$, as measured by its effect on a probe scalar field,
stays a constant over the entire space-time.

%Notice that the state ${\cal O}\vert {\rm phys}\rangle$ is also a
%physical state as long as ${\cal O}$ is BRST invariant. 
%The gauge invariance is just a restricted form of the BRST invariance
%to the case that the operator ${\cal O}$ is solely written 
%in term of $h^{\mu\nu}$ without containing $c^\mu$, $\bar c^\mu$ and $B^\mu$. 
%Thus, we conclude  $\langle{\rm phys}\vert {\cal O}[g_{\mu\nu} \tilde
%G^{\mu\nu}-(-R)]\vert{\rm phys}\rangle$=0. 

\section{conclusion}

A secular screening of the cosmological constant by infrared
quantum effects would represent a very spectacular phenomenon in
low energy quantum gravity. In this note, we have reanalyzed the
issue of gauge invariance in the definition of the expansion rate
$H(t)$ which was used in the original analysis of this 
problem~\cite{tw} (see Eq.~(\ref{H})).

We have shown that such definition is only invariant under c-number
gauge transformations, but not under generic changes of the gauge
fixing term. Such changes correspond to gauge transformations
where the gauge parameter $\chi^{\mu}$ depends on the operator
$h_{\mu\nu}$. Because of that, they introduce arbitrary time
dependence in the expansion rate $H(t)$ as defined in 
Eq.~(\ref{H}). Hence, the interpretation of the results in Ref.~\cite{tw} 
as a physical screening of $\Lambda$ seems very
questionable.

A truly gauge invariant definition of $H(t)$ was introduced in
Ref.~\cite{Abramo:2001db}. 
This definition was motivated on physical grounds
as follows. A constant source $J$ in a quasi-de Sitter universe
coupled to a conformal scalar field $\phi$ will produce a field
$\phi(t)$. The amplitude of a free conformal scalar in quasi-de
Sitter decays with time like the inverse of the scale factor.
Hence, the late time behaviour of $\phi(t)$ is dominated by the
contribution of the source during the last e-folding time, and is
therefore proportional to the surface of a sphere of Hubble size
$$\phi(t)\propto J H^{-2}(t)~.$$ 
It was proposed in~\cite{Abramo:2001db} 
that such auxiliary field be used as a measure of the
local expansion rate. The field $\phi$ is given by
the inverse of the perturbed wave operator acting on the constant source. 
This is a non-local and rather cumbersome expression to deal with in the quantum theory. 
On the other hand, we have argued that there is an
alternative definition which is equally useful if we wish to monitor
an adiabatic change in the expansion rate (such as the one which would 
be suggested by Eq.~(\ref{decay})). Indeed the curvature scalar ${\cal R}$ is proportional to $H^2(t)$ 
plus corrections of order $\,\dot H$ which are negligible
in the adiabatic limit. So the question  is whether the value of this scalar 
(or a suitable smearing of it) can
change in the course of time. Classically, for the system of pure gravity coupled to a
cosmological constant, this is impossible. By using the path integral approach, 
we confirmed that this conclusion is not altered when we take into account 
the subtleties associated with gauge invariance and renormalization. 
Therefore, according to this definition,
we find no evidence of a secular screening of the cosmological
constant, to all orders in perturbation theory.

It should be stressed that these arguments apply only to the case
of pure gravity with a cosmological constant, and they do not
exclude the possibility of interesting infrared effects in
theories with a different field 
content~\cite{Abramo:2001dc,Vilenkin:1982wt,Losic:2005vg,Perez} or due to
non-perturbative effects~\cite{Antoniadis:2006wq}.
Our considerations focused on the renormalized expectation value
of the scalar curvature ${\cal R}$, which is insensitive to the 
classical back-reaction effect due to a bath of gravitons. In future work, we 
would like to examine different gauge invariant indicators of the expansion 
rate~\cite{Tsamis:2005bh}, 
which give a non-vanishing result depending on the choice of the initial state.\\

\noindent
{\bf Note added:}\\

After this paper was submitted to the archives, Tsamis and Woodard
wrote a reply to it~\cite{TWR}, expressing some points of view which we
do not share.

First, they claim that we did not show that the
renormalized Ricci scalar is constant, and that our Eq.~(\ref{lt2}) is completely 
consistent with screening. The observable we 
calculate is the expectation value of the integral of the Ricci
scalar over a region of space-time. { This operator is divergent, and so
we define the corresponding renormalized operator by standard techniques.} 
We show that this agrees with the expectation value 
of the integral of a constant, over the same region. The equality holds order by order
in the loop expansion. The region of space-time is itself arbitrary, as long as the same one is used
on both sides of the equation. { In our view, this means that the renormalized Ricci scalar, as measured by
its effect on a probe scalar field, stays constant, in as precise a sense as can be made.} 
Notice that this is exactly the condition for $J$ to be constant 
with constant $\phi$ in Eq.~(\ref{source}). 

The authors of Ref.~\cite{TWR} object that we use an external scalar window function $W(x)$ in our
definition of the gauge invariant operator { (The reason for that is explained in our
footnote~\ref{footnote6}). This scalar is not constructed from the metric, and hence, the integral 
of $\sqrt{g}W(x){\cal R}$ does not correspond to any
observable of the theory: it depends on the particular choice of $W(x)$. 
However the statement that our equality (\ref{lt2}) holds for any $W(x)$ is, of course, independent 
of this choice, and hence it is a physically meaningful statement.}
They also object that even if we
show that  $\langle {\cal R}^{\rm ren} \rangle_W=4 \Lambda\ \langle 1^{\rm
ren}\rangle_W\ $, both sides of the equation can evolve secularly in the same way. 
Even if that were the case, this would not imply any secular evolution of their 
ratio, {which is the quantity of our interest (for constant $\phi$, the ratio is proportional to $J$ at 
the classical level).}
They also claim, at the beginning of Section 3 
that our renormalization scheme is "peculiar". We disagree with this appreciation. 
What we do is standard renormalization in low energy effective theory. We do make 
a particular choice for the finite parts of the local counterterms which need to be 
subtracted. { This is explained in detail in the paragraphs between our Eqs.~(\ref{lt2}) 
and (\ref{phi=0})}. The important point is that there is a choice of counterterms for 
which there is no secular screening of the renormalized operator. If a 
change in the local counterterms happened to give rise to some additional effect, 
then this would be an effect due to local physics
 (or, conceivably, to the secular 
evolution of the added higher-order local counter terms, although this seems unlikely), 
but it would be unrelated to the infrared secular evolution of ${\cal R}$.

The authors of~\cite{TWR} also purport that if we are allowed arbitrary
subtractions in order to construct the renormalized operator ${\cal R}_{\rm ren}$, then
we could absorb in its definition things like the one loop effective 
potential of a scalar field. If so, they argue, we would reach the conclusion that 
${\cal R}_{\rm ren}$ stays constant even in a theory like "new inflation", where 
the potential is due to one loop corrections. Of course this would not be correct,
and it has nothing to do with the method we are using in the present paper. 
Arbitrary subtractions are simply not allowed. At each order in the loop expansion, 
we only allow as counterterms a finite number of higher dimension local operators, 
suppressed by corresponding powers of the Planck mass $M_{pl}$. 
The number of counterterms will be larger if we work at a higher order, because this is 
unavoidable in non-renormalizable theories. 
But these higher order counterterms can never absorb the lower order
loop-corrections since the power of $M_{pl}$ is different. 
%non-local 
%contributions (or infrared contributions to the effective potential in the case of a theory 
%with matter scalar fields).

We would agree that there are other observables one can look at. 
Our claim is that we see no evidence for a 
secular infrared screening in the observable we have analyzed. We should add that this is 
a better defined observable than the spatially averaged Hubble rate used in~\cite{tw}. 
The authors of~\cite{TWR} claim in Section 2 of their reply that gauge dependent 
quantities can have some physical content. While this is debatable, 
their discussion does not seem to 
warrant the preference of a gauge-dependent result over the gauge-invariant one we
presented in this paper.

\section{Acknowledgments}

J.G. is grateful to Enric Verdaguer and Alex Pomarol for interesting discussions.
T.T. would like to thank Jiro Soda and Misao Sasaki for their valuable
comments. {We also thank Alexei Starobinsky, Nicholas Tsamis, Richard Woodard and Bojan
Losic for valuable comments.} 
The work of J.G. is partially supported by CICYT grant FPA
2004-04582-C02-02 and DURSI 2001-SGR-0061. 
%M.P. is supported in part by NSF grant PHY-0245068. 
The work of T.T. is partially supported by 
Monbukagakusho Grant-in-Aid
for Scientific Research Nos.~17340075 and~19540285, %kiban(B)
and also in part by the 21st Century COE 
``Center for Diversity and Universality in Physics'' at Kyoto
 university, from the Ministry of Education,
Culture, Sports, Science and Technology of Japan.


\begin{thebibliography}{99}

\bibitem{staro} 
A.~A.~Starobinsky, JETP\ Lett.\ {\bf 30}, 682 (1979).

\bibitem{tw} 
%Tsamis and Woodard
%\bibitem{Tsamis:1996qq}
  N.~C.~Tsamis and R.~P.~Woodard,
  %``Quantum Gravity Slows Inflation,''
  Nucl.\ Phys.\  B {\bf 474}, 235 (1996);
%  [arXiv:hep-ph/9602315].
  %%CITATION = NUPHA,B474,235;%%
%\bibitem{Tsamis:1996qm}
%  N.~C.~Tsamis and R.~P.~Woodard,
  %``The quantum gravitational back-reaction on inflation,''
  Annals Phys.\  {\bf 253}, 1 (1997).
%  [arXiv:hep-ph/9602316].
  %%CITATION = APNYA,253,1;%%

\bibitem{wo04}
  R.~P.~Woodard,
  %``de Sitter breaking in field theory,''
  arXiv:gr-qc/0408002.
  %%CITATION = GR-QC/0408002;%%

%\bibitem{Weinberg:2000yb}
%  S.~Weinberg,
%  %``The cosmological constant problems,''
%  arXiv:astro-ph/0005265.
%  %%CITATION = ASTRO-PH/0005265;%%

\bibitem{Abramo:2001db}
  L.~R.~Abramo and R.~P.~Woodard,
  %``A Scalar Measure Of The Local Expansion Rate,''
  Phys.\ Rev.\  D {\bf 65}, 043507 (2002)
  [arXiv:astro-ph/0109271].
  %%CITATION = PHRVA,D65,043507;%%

\bibitem{unruh98}
  Related arguments can be found in W.~Unruh,
  %``Cosmological long wavelength perturbations,''
  arXiv:astro-ph/9802323. There, the same expansion parameter
is used for the metric perturbations and for the change in the gauge. 
Here we are considering an infinitessimal gauge transformation, 
while the metric perturbation is not necessarily small.

  %%CITATION = ASTRO-PH/9802323;%%

\bibitem{MTW}
 C.~W.~Misner, K.~S.~Thorne, and J.~A.~Wheeler, {\it Gravitation}, 
 (W.H. Freeman \& Co., San Francisco, 1973).


\bibitem{Abramo:2001dc}
  L.~R.~Abramo and R.~P.~Woodard,
  %``No one loop back-reaction in chaotic inflation,''
  Phys.\ Rev.\  D {\bf 65}, 063515 (2002)
  [arXiv:astro-ph/0109272].
  %%CITATION = PHRVA,D65,063515;%%

\bibitem{Vilenkin:1982wt}
  See e.g. A.~Vilenkin and L.~H.~Ford,
  %``Gravitational Effects Upon Cosmological Phase Transitions,''
  Phys.\ Rev.\  D {\bf 26}, 1231 (1982).
 %%CITATION = PHRVA,D26,1231;%%

\bibitem{Losic:2005vg}
  B.~Losic and W.~G.~Unruh,
  %``Long-wavelength metric backreactions in slow-roll inflation,''
  Phys.\ Rev.\  D {\bf 72}, 123510 (2005)
  [arXiv:gr-qc/0510078].
  %%CITATION = PHRVA,D72,123510;%%


\bibitem{kugo}
  T.~Kugo and S.~Uehara,
  %``General Procedure Of Gauge Fixing Based On Brs Invariance Principle,''
  Nucl.\ Phys.\  B {\bf 197}, 378 (1982).
  %%CITATION = NUPHA,B197,378;%%

\bibitem{Perez} G. Perez-Nadal, A. Roura and E. Verdaguer, in preparation.

\bibitem{Antoniadis:2006wq} For a discussion of possible non-perturbative effects 
leading to a conformal phase of gravity, see
  I.~Antoniadis, P.~O.~Mazur and E.~Mottola,
  %``Cosmological dark energy: Prospects for a dynamical theory,''
  New J.\ Phys.\  {\bf 9}, 11 (2007)
  [arXiv:gr-qc/0612068], and references therein.
  %%CITATION = NJOPF,9,11;%%

\bibitem{Losic:2006ht}
  B.~Losic and W.~G.~Unruh,
  %``On leading order gravitational backreactions in de Sitter spacetime,''
  Phys.\ Rev.\  D {\bf 74}, 023511 (2006)
  [arXiv:gr-qc/0604122].
  %%CITATION = PHRVA,D74,023511;%%

\bibitem{Hollands:2004yh}
  S.~Hollands and R.~M.~Wald,
  %``Conservation of the stress tensor in interacting quantum field theory  in
  %curved spacetimes,''
  Rev.\ Math.\ Phys.\  {\bf 17}, 227 (2005)
  [arXiv:gr-qc/0404074].
  %%CITATION = RMPHE,17,227;%%

%\cite{Tsamis:2005bh}
\bibitem{Tsamis:2005bh}
  e.g. N.~C.~Tsamis and R.~P.~Woodard,
  %``A measure of cosmological acceleration,''
  Class.\ Quant.\ Grav.\  {\bf 22}, 4171 (2005).
%  [arXiv:gr-qc/0506089].
  %%CITATION = CQGRD,22,4171;%%


\bibitem{TWR}
  N.~C.~Tsamis and R.~P.~Woodard,
  %``Reply to `Can infrared gravitons screen $\Lambda$?',''
  arXiv:0708.2004 [hep-th].


\end{thebibliography}
\end{document}